\documentclass[11pt,a4paper]{article}
\usepackage[latin1]{inputenc}
\usepackage{pdflscape}

\usepackage{graphicx,booktabs}
\usepackage{multicol,wrapfig}

\usepackage{amsfonts}
\usepackage{amsmath}
\usepackage{amssymb}
\usepackage{theorem}
\usepackage{upgreek}
\usepackage{xcolor}
\usepackage{float}
\usepackage{rotating}

\theoremstyle{change}
\theorembodyfont{\slshape}
\newtheorem{satz}{Theorem}[section]

\newtheorem{prop}[satz]{Proposition}

\theorembodyfont{\upshape\small}

\theorembodyfont{\ttfamily}

\newcommand{\ba}{\begin{equation}}
\newcommand{\ea}{\end{equation}}

\usepackage{dsfont}
\newcommand{\indfkt}{\mathds{1}}

\newcommand{\fp}{\mbox{\boldmath $p$}}

\newcommand{\fX}{\mbox{\boldmath $X$}}
\newcommand{\fx}{\mbox{\boldmath $x$}}

\newcommand{\fY}{\mbox{\boldmath $Y$}}

\newcommand{\fZ}{\mbox{\boldmath $Z$}}

\newcommand{\mT}{\mbox{\textup{\textbf{T}}}}

\newcommand{\expon}{\textup{Exp}}
\newcommand{\norm}{\textup{N}}

\newcommand{\bbn}{\mathbb{N}}

\newcommand{\iid}{i.\,i.\,d.}
\newcommand{\ie}{i.\,e., }
\newcommand{\eg}{e.\,g., }

\usepackage{natbib}

\usepackage{hyperref}

\begin{document}



\parindent 0cm

\title{Using Transcripts for Nonparametric Monitoring of Serial Dependence}

\author{
Christian H.\ Wei\ss{}\thanks{Department of Mathematics and Statistics, Helmut Schmidt University, 22043 Hamburg, Germany}\ \thanks{Corresponding author. E-Mail: \href{mailto:weissc@hsu-hh.de}{\nolinkurl{weissc@hsu-hh.de}}. ORCID: \href{https://orcid.org/0000-0001-8739-6631}{\nolinkurl{0000-0001-8739-6631}}.}
\and 
Jos\'e M.\ Amig\'o\thanks{Centro de Investigaci\'on Operativa, Universidad Miguel Hern\'andez, 03202 Elche, Spain}\ \thanks{E-Mail: \href{mailto:jm.amigo@umh.es}{\nolinkurl{jm.amigo@umh.es}}. ORCID: \href{https://orcid.org/0000-0002-1642-1171}{\nolinkurl{0000-0002-1642-1171}}.}
}

\maketitle

\begin{abstract}
Control charts for process monitoring are widely used in practice. Most control charts require the monitored (residuals) process to be serially independent (and to satisfy specified distributional assumptions), whereas undetected dependence (or violations of distributional assumptions) may severely affect the charts' performances. Therefore, (distribution-free) control charts for monitoring serial dependence are of utmost relevance for practice. 
Recently, various nonparametric control charts have been proposed for this purpose, which are based on ordinal patterns, and which showed an appealing performance in detecting different types of serial dependence. In this research, we further progress in this direction and develop novel nonparametric control charts being based on transcripts and algebraic distances (as derived from ordinal patterns). The performance of the newly proposed control charts is evaluated in a simulation study, and their application in practice is illustrated with a real-world data example from chemical industry.

\end{abstract}


\section{Introduction}
\label{Introduction}

Since their proposal by Walter A.\ Shewhart more than 100~years ago (see \citealp{olmstead67} for historical aspects), control charts received an enormous research interest and have been widely applied in practice in order to monitor a process $(X_{t})=(X_{t})_{t\in \bbn=\{1,2,\ldots \}}$ for possible process changes; see \citet{mont09} for an overview. More precisely, a control chart can be understood as a sequential hypothesis test, with the in-control (IC) model constituting the null hypothesis. If the process $(X_{t})$ experiences a change at some time $\tau\in\bbn$ (referred to as the change point), 
then it turns into an out-of-control (OOC) state (alternative hypothesis), the timely detection of which being the aim of statistical process monitoring (SPM). The vast majority of control charts that have been proposed so far (see \citet{mont09} for references) assume that the monitored process is serially independent under IC-conditions; see \citet{chakraborti20} for a discussion. However, if this assumption is violated, it is well known since \citet{alwan92} and \citet{wardell92} that the control chart's performance might be severely affected. Therefore, having a control chart that monitors the IC-assumption of serial dependence is crucial for practice. An analogous conclusion applies to the case of time-series monitoring, where control charts are typically applied to a residual series, which is serially uncorrelated under IC-conditions \citep{yourstone89,atienza97}. 

\smallskip
In real-world applications, the monitored process $(X_{t})$ often does not follow a standard parametric distribution (such as the normal distribution). Hence, if the chart design relies on inappropriate distributional assumptions, the chart may exhibit an unexpected performance (such as frequent false alarms under IC-conditions or detection delays under OOC-conditions); see \citet{chakraborti20} for a discussion. This motivates the use of nonparametric (distribution-free) control charts as recently surveyed by \citet{chakraborti19} and \citet{koutras20}, which show the same performance irrespective of the actual marginal distribution of $(X_{t})$. Recently, \citet{weiss23} developed a bunch of nonparametric control charts that monitor for serial dependence in $(X_t)$. While these charts are designed for continuously distributed processes $(X_t)$, with arbitrary marginal distribution and being independent and identically distributed (\iid)\ under IC-conditions, they can also be applied to discrete-valued processes (such as a count process) if additional jittering \citep{machado05} is used (\eg if \iid\ uniform noise with range $(0;1)$ is added to a count process, see \citealp{weiss24}). Thus, without loss of generality, let us focus on continuously distributed processes $(X_t)$ in the sequel, which are assumed to be \iid\ under IC-conditions. 

\smallskip
The nonparametric control charts for serial dependence developed by \citet{weiss23} combine so-called ordinal patterns (OPs) with an exponentially weighted moving-average (EWMA) approach; see Section~\ref{Review: Control Charts based on Ordinal Patterns} for a detailed review. 
In a recent research on nonparametric hypothesis tests for serial dependence by \citet{weiss26}, however, it was shown that superior power is often achieved if the test statistics do not rely on ordinary OPs, but if so-called transcripts (and possibly related algebraic distances) are considered instead. Therefore, in the present research, we develop novel nonparametric control charts being based on the transcripts computed from $(X_t)$, and we compare their performance to the former OP-based control charts of \citet{weiss23}. The relevant background on transcripts and algebraic distances is provided by Section~\ref{Review: Transcripts and Algebraic Distances}, and our novel transcript-based control charts are introduced in Section~\ref{Novel Control Charts based on Transcripts}. Their performance, evaluated in terms of the average run length (ARL) under IC- or OOC-conditions, is investigated by simulations in Section~\ref{Performance Analyses}, while an illustrative data example from chemical industry is presented in Section~\ref{Illustrative Data Application}. A summary, the main conclusions, and an outlook are the contents of the final Section~\ref{Conclusions}.


\section{Review: Control Charts based on Ordinal Patterns}
\label{Review: Control Charts based on Ordinal Patterns}
The idea to compute OPs from a real-valued and continuously distributed process $(X_{t})$ dates back to \citet{bandt02}. Let $m\in \mathbb{N}=\{1,2,\ldots \}$ with $m\geq 2$ be the length of the considered OPs, and let $S_{m}$ denote the symmetric group of degree~$m$, which consists of the $m!$ possible permutations of the integers $\{1,\ldots ,m\}$ endowed with function composition. The permutations~$\pi \in S_{m} = \{\pi^{[1]}, \ldots, \pi^{[m!]}\}$ can be used in various ways to represent the $m!$ different OPs of a vector $\fx=(x_{1},\ldots ,x_{m})\in \mathbb{R}^{m}$, see \citet{berger19} for a discussion. In what follows, we focus on the \emph{permutation representation}, where $\pi =(i_{1},\ldots ,i_{m})\in S_{m}$ expresses that permutation, which causes an ascending order of the components of~$\fx$: 
\begin{equation}
x_{i_{1}}\leq x_{i_{2}}\leq \ldots \leq x_{i_{m}},\quad \text{and}\quad
i_{k-1}<i_{k}\text{ if }x_{i_{k-1}}=x_{i_{k}}\text{ for }k\geq 2.
\label{ordpattern}
\end{equation}
The second case in \eqref{ordpattern} accounts for the possible occurrence of ties in~$\fx$. As $(X_t)$ is continuously distributed, ties happen with probability zero. However, due to the limited measurement accuracy in real-world applications, ties may happen anyway, but we assume that they occur at an negligible rate. 

\smallskip
By continuously mapping the segments $\fX_{t}=(X_{t},\ldots,X_{t+m-1})$ of the process $(X_{t})$ onto its corresponding OP~$\pi _{t}$, for $t=1,2,\ldots$, the originally real-valued process $(X_{t})$ is discretized and transformed into the OP-series $(\pi _{t})=(\pi _{t})_{t\in \mathbb{N}}$ in an online manner. According to the proposal by \citet{weiss23}, the process $(X_{t})$ is now monitored by monitoring its corresponding OP-series $(\pi _{t})$.
If the original process $(X_t)$ runs under IC-conditions, \ie if it is \iid, then the OP-series $(\pi _{t})$ is stationary with a discrete-uniform marginal distribution, irrespective of the actual marginal distribution of $(X_t)$. Hence, its $m!$-dimensional probability mass function (PMF) vector $\fp_\pi = (p_{\pi,1}\ldots, p_{\pi,m!})^\top$ with $p_{\pi,k} = P(\pi_t=\pi^{[k]})$ is equal to $\fp_{\pi}{}^{(0)}=(1/m!,\ldots,1/m!)$. Deviations between~$\fp_\pi$ and~$\fp_{\pi}{}^{(0)}$, in turn, indicate the presence of serial dependence in $(X_t)$. 

\smallskip
More precisely, \citet{weiss23} solely focused on OPs of length $m=3$, which is indeed the most common choice in practice, see \citet{bandt19}. Then, the six possible OPs (in lexicographic order) can be summarized as follows:
\ba
\label{OrdPatt3}
\begin{array}{@{}lcccccc@{}}
\text{Perm.:} & (1,2,3) & (1,3,2) & (2,1,3) & (2,3,1) & (3,1,2) & (3,2,1) \\
\text{Obs.:}\\[-3ex]
 & \includegraphics[viewport=75 85 135 145, clip=, scale=0.65]{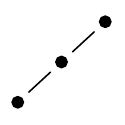}
 & \includegraphics[viewport=75 85 135 145, clip=, scale=0.65]{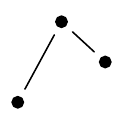} 
 & \includegraphics[viewport=75 85 135 145, clip=, scale=0.65]{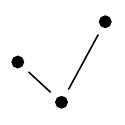} 
 & \includegraphics[viewport=75 85 135 145, clip=, scale=0.65]{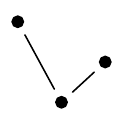} 
 & \includegraphics[viewport=75 85 135 145, clip=, scale=0.65]{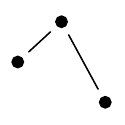} 
 & \includegraphics[viewport=75 85 135 145, clip=, scale=0.65]{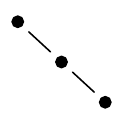} 
\end{array}
\ea
For process monitoring, the online estimation of the OP series' current PMF vector is necessary. \citet{weiss23} propose an EWMA approach for this purpose, namely by computing
\ba
\label{EWMA}
\hat{\fp}_{\pi,0}\ =\ \fp_{\pi}^{(0)},\qquad
\hat{\fp}_{\pi,t}\ =\ \lambda\,\fY_t + (1-\lambda)\,\hat{\fp}_{\pi,t-1} \quad\text{for } t=1,2,\ldots
\ea
Here, the binary vectors~$\fY_t{}$ of dimension $m!=6$ are defined by the ``one-hot encoding'' $Y_{t,k}{}=\indfkt(\pi_t{}=\pi^{[k]})$. The smoothing parameter $\lambda\in (0;1)$ is chosen by the user, where the extent of smoothing and, thus, the inherent memory increases for decreasing~$\lambda$. At each time~$t$, the EWMA estimate~$\hat{\fp}_{\pi,t}$ is used to compute the value of the considered monitoring statistic, which is then plotted on a control chart and compared against the actual control limits (CLs). 

\smallskip
In their research, \citet{weiss23} compared six versions of such monitoring statistics. The three kinds of entropy-like statistics and resulting charts showed virtually the same performance, so it is sufficient to present only one of them here. As its statistic is most easily computed, we focus on the $\Delta_\pi$-chart in the sequel, which plots the statistics
\ba
\label{Delta_pi-chart}
\widehat{\Delta}_{\pi,t}\ =\ \sum_{k=1}^{m!} \big(\hat{p}_{\pi,t,k}-1/m!\big)^2
\quad\text{for } t=1,2,\ldots
\ea
against an upper CL (UCL) $l_{\Delta_\pi}>0$. Among the remaining three charts in \citet{weiss23}, only the $\uptau$-chart and the $\upbeta$-chart showed an appealing performance. The $\uptau$-chart plots the statistics
\ba
\label{tau-chart}
\widehat{\uptau}_t\ =\ \hat{p}_{\pi,t,1}+\hat{p}_{\pi,t,6}-1/3
\quad\text{for } t=1,2,\ldots
\ea
against an UCL $l_{\uptau}>0$ and the corresponding lower CL (LCL) $-l_{\uptau}$ (\ie symmetric two-sided CLs), and the $\upbeta$-chart the statistics
\ba
\label{beta-chart}
\widehat{\upbeta}_t\ =\ \hat{p}_{\pi,t,1}-\hat{p}_{\pi,t,6}
\quad\text{for } t=1,2,\ldots
\ea
against an UCL $l_{\upbeta}>0$ and the corresponding LCL $-l_{\upbeta}$. These three control charts \eqref{Delta_pi-chart}--\eqref{beta-chart} shall serve as the competitors for our newly proposed control charts in Section~\ref{Novel Control Charts based on Transcripts} below. Since the charts are nonparametric, the actual values of the CL parameters $l_{\Delta_\pi},l_{\uptau},l_{\upbeta}$ are unique across different marginal distributions of $(X_t)$, where the CL~values for $\lambda\in\{0.25,0.10,0.05\}$ are summarized in Table~1 of \citet{weiss23}.


\section{Review: Transcripts and Algebraic Distances}
\label{Review: Transcripts and Algebraic Distances}

As said in Section \ref{Review: Control Charts based on Ordinal Patterns}, the permutations of $m\geq 2$ elements $\{1,2,\ldots,m\}$ (without restriction), endowed with function composition \textquotedblleft $\circ $\textquotedblright , build a group called the symmetric
group of degree $m$ and denoted $S_{m}$. Among several possibilities, here we choose the one-line notation $(i_{1},i_{2},\ldots ,i_{m})$ for the permutation $n\mapsto i_{n}$ ($1\leq n,i_{n}\leq m$); this notation will allow us below to leverage edit
distances between permutations (viewed as strings of non-repeated symbols).
Following the usual convention in the literature, the composition $\pi
_{1}\circ \pi _{2}$ of the permutations (or OPs for that matter) $\pi
_{1}=(i_{1},i_{2},\ldots ,i_{m})$ and $\pi _{2}=(j_{1},j_{2},\ldots ,j_{m})$
is defined as a \textquotedblleft right action\textquotedblright , meaning that the permutation $\pi _{1}$ acts first and $\pi _{2}$ acts second, \ie 
\begin{equation}
\pi _{1}\circ \pi _{2}=(i_{1},i_{2},\ldots ,i_{m})\circ (j_{1},j_{2},\ldots
,j_{m})=(j_{i_{1}},j_{i_{2}},\ldots ,j_{i_{m}}).  \label{pi1 x p2}
\end{equation}%
To harness the algebraic structure of $S_{m}$ in the analysis of time series
symbolized with OPs, \citet{monetti09} introduced the concept of \emph{%
transcript} between OPs. Specifically, given two OPs $\pi
_{1},\pi _{2}\in S_{m}$, the transcript from the source $\pi _{1}$ to the
target $\pi _{2}$ is the OP $\tau\in S_{m}$
defined as%
\begin{equation}
\tau =\tau (\pi _{1},\pi _{2}):=\pi _{2}\circ \pi _{1}^{-1}.
\label{transcrip}
\end{equation}%
The order of the source and target permutations matters because
\begin{equation}
\tau (\pi _{2},\pi _{1})=\pi _{1}\circ \pi _{2}^{-1}=\tau (\pi _{1},\pi
_{2})^{-1}\text{.}  \label{transcript2}
\end{equation}

For other properties of the transcripts, see \citet{amigo12}.

\smallskip
As way of illustration, the composition (multiplication or Cayley) table $(\pi _{i}\circ \pi _{j})_{1 \leq i,j \leq m}$ of the group $S_{3}$ is given by (see equation \eqref{pi1 x p2} with $m=3$): 

\ba
\label{tabOP}
\begin{array}{c|cccccc}
\toprule
\text{Composition } \circ & \pi^{[1]} & \pi^{[2]} & \pi^{[3]} & \pi^{[4]} & \pi^{[5]} & \pi^{[6]} \\
\midrule
\pi^{[1]} = (1,2,3) & \pi^{[1]} & \pi^{[2]} & \pi^{[3]} & \pi^{[4]} & \pi^{[5]} & \pi^{[6]} \\
\pi^{[2]} = (1,3,2) & \pi^{[2]} & \pi^{[1]} & \pi^{[4]} & \pi^{[3]} & \pi^{[6]} & \pi^{[5]} \\
\pi^{[3]} = (2,1,3) & \pi^{[3]} & \pi^{[5]} & \pi^{[1]} & \pi^{[6]} & \pi^{[2]} & \pi^{[4]} \\
\pi^{[4]} = (2,3,1) & \pi^{[4]} & \pi^{[6]} & \pi^{[2]} & \pi^{[5]} & \pi^{[1]} & \pi^{[3]} \\
\pi^{[5]} = (3,1,2) & \pi^{[5]} & \pi^{[3]} & \pi^{[6]} & \pi^{[1]} & \pi^{[4]} & \pi^{[2]} \\
\pi^{[6]} = (3,2,1) & \pi^{[6]} & \pi^{[4]} & \pi^{[5]} & \pi^{[2]} & \pi^{[3]} & \pi^{[1]} \\
\bottomrule
\end{array}
\ea

Hence, the transcripts $\tau (\pi _{1},\pi _{2})=\pi _{2}\circ \pi _{1}^{-1}$
for $\pi _{1},\pi _{2}\in S_{3}$ (see equation \eqref{transcrip}) are given
in the following table, where the source OP $\pi _{1}$ labels the rows
(leftmost column) and the target OP $\pi _{2}$ labels the columns (topmost
row):


\ba
\label{tab_tr}
\begin{array}{c|cccccc}
\toprule
\text{Transcript } \tau & \pi^{[1]} & \pi^{[2]} & \pi^{[3]} & \pi^{[4]} & \pi^{[5]} & \pi^{[6]} \\
\midrule
\pi^{[1]} = (1,2,3) & \pi^{[1]} & \pi^{[2]} & \pi^{[3]} & \pi^{[4]} & \pi^{[5]} & \pi^{[6]} \\
\pi^{[2]} = (1,3,2) & \pi^{[2]} & \pi^{[1]} & \pi^{[5]} & \pi^{[6]} & \pi^{[3]} & \pi^{[4]} \\
\pi^{[3]} = (2,1,3) & \pi^{[3]} & \pi^{[4]} & \pi^{[1]} & \pi^{[2]} & \pi^{[6]} & \pi^{[5]} \\
\pi^{[4]} = (2,3,1) & \pi^{[5]} & \pi^{[6]} & \pi^{[2]} & \pi^{[1]} & \pi^{[4]} & \pi^{[3]} \\
\pi^{[5]} = (3,1,2) & \pi^{[4]} & \pi^{[3]} & \pi^{[6]} & \pi^{[5]} & \pi^{[1]} & \pi^{[2]} \\
\pi^{[6]} = (3,2,1) & \pi^{[6]} & \pi^{[5]} & \pi^{[4]} & \pi^{[3]} & \pi^{[2]} & \pi^{[1]} \\
\bottomrule
\end{array}
\ea

\medskip
From equation \eqref{transcrip}, we conclude that the transcript $\tau (\pi
_{1},\pi _{2})$ is the permutation $\tau $ that transforms~$\pi _{1}$ into~$%
\pi _{2}$, in the sense that $\tau \circ \pi _{1}=\pi _{2}$. Thus, it expresses a kind of ``dissimilarity'' between~$\pi_{1}$ and~$\pi_{2}$, where~$\pi^{[1]}$ corresponds to least dissimilarity according to \eqref{tab_tr}. Therefore, it
is not surprising that transcripts are related to the following two types of
algebraic distance between two permutations \citep{amigo25}:

\begin{description}
\item(1) The \emph{Cayley distance} $d_{\textup{C}}:S_{m}\times
S_{m}\rightarrow \{0,1,\ldots ,m-1\}$ between $\pi _{1},\pi _{2}\in S_{m}$
is defined as the minimum number of transpositions needed to transform~$\pi
_{1}$ into~$\pi _{2}$. For example, $d_{\textup{C}}((1,2,3),(3,2,1))=1$. It can be shown that%
\begin{equation}
d_{\textup{C}}(\pi _{1},\pi _{2})=m-C(\tau (\pi _{1},\pi _{2})),
\label{D_C}
\end{equation}%
where $C(\tau (\pi _{1},\pi _{2}))$ is the number of cycles (including $1$-cycles) in the cycle
factorization of the transcript $\tau =\pi _{2} \circ \pi _{1}^{-1}$ \citep{nguyen24}.

\item(2) The \emph{Kendall distance} $d_{\textup{K}}:S_{m}\times
S_{m}\rightarrow \{0,1,\ldots m(m-1)/2\}$ between $\pi _{1},\pi _{2}\in S_{m}
$ is defined as the minimum number of adjacent transpositions needed to
transform~$\pi _{1}$ into~$\pi _{2}$. For example, $d_{\textup{K}}((1,2,3),(3,2,1))=3$. It can be shown that%
\begin{equation}
d_{\textup{K}}(\pi _{1},\pi _{2})=I(\tau (\pi _{1},\pi _{2})),  \label{d_K}
\end{equation}%
where $I(\tau (\pi _{1},\pi _{2}))$ is the number of inversions in the transcript $\tau =\pi _{2} \circ \pi _{1}^{-1}$, \ie the number of ordered pairs $(k_{i},k_{j})$ in $\tau
=(k_{1},\ldots,k_{m})$ such that $k_{i}>k_{j}$ \citep{kendall1938}. 
\end{description}

Equations \eqref{D_C} and \eqref{d_K} are used to compute the respective distances in practice. Note that (i) $d_{\textup{C}}(\pi _{1},\pi _{2}) \leq d_{\textup{K}}(\pi _{1},\pi _{2})$ for all $\pi _{1},\pi _{2} \in S_{m}$, and (ii) the range of $d_{\textup{K}}$ is larger than
the range of $d_{\textup{C}}$ except for $m=2$ (in
which case both ranges are $\{0,1\}$). Therefore, we expect $d_{K}$ to have more discrimination power in time
series analysis than $d_{C}$. For other transcript-based tools in time series analysis, see \citet{amigo26}.

\medskip
In the present research, we particularize the length of the OPs (and, hence, of the transcripts) and the algebraic distance, as follows.

\begin{itemize}
\item The length of the OPs is $m=3$ from now on. This is not only the most
common length used in applications in general \citep{bandt19}, and for the control charts of \citet{weiss23} in particular, recall Section~\ref{Review: Control Charts based on Ordinal Patterns}, but it is also the
length used in \citet{weiss26}, some of whose results are
instrumental for the present work. See \eqref{tabOP} and \eqref{tab_tr} for the corresponding multiplication and transcript's tables.

\item \citet{weiss26} also showed numerically that the Kendall distance $%
d_{\textup{K}}$ outperforms the Cayley distance $d_{\textup{C}}$ when it comes to detecting
serial dependence in time series analysis. For this reason, we only
use $d_{\textup{K}}$ for monitoring (in addition to the transcript series itself).
\end{itemize}

Thus, let $(X_{t})$ be a real-valued and continuously distributed process
having a stationary OP-series $(\pi _{t})$ of length $m=3$. The
corresponding transcript series $(\tau _{t})$ is defined by $\tau _{t}=\tau(\pi _{t},\pi _{t+1}):=\pi
_{t+1}\circ \pi _{t}^{-1}$ according to \eqref{transcrip}, and let $(d_{\textup{K},t})$ be the resulting
series of Kendall distances, that is, $d_{\textup{K},t}=d_{\textup{K}}(\pi _{t},\pi _{t+1})=I(\tau _{t})$, see equation \eqref{d_K}.

\medskip
It should be noted, however, that for two consecutive OPs such as $\pi_{t}$ and $\pi_{t+1}$ in $(\tau_t)$, some transitions are impossible, irrespective of the underlying process $(X_t)$. As summarized in Figure~1 of \citet{sousa22}, it is impossible that $\pi^{[k]}$ with $k\in\{1,3,4\}$ is followed by $\pi^{[l]}$ with $l\in\{3,4,6\}$, and analogously with $k\in\{2,5,6\}$ and $l\in\{1,2,5\}$, where we took into account that \citet{sousa22} use different rules than ours to define and order the OPs. 
Altogether, each transcript $\tau_{t}$ corresponds to a set of possible 4-OPs of the vector $(x_t,x_{t+1},x_{t+2},x_{t+3})$, where $\tau_{t}=\pi
_{t+1}\circ \pi _{t}^{-1}$, $\pi _{t}$ is the 3-OP of $(x_t,x_{t+1},x_{t+2})$, and  $\pi _{t+1}$ is the 3-OP of $(x_{t+1},x_{t+2},x_{t+3})$; see Table~\ref{tabTrLag0} for a summary.

\begin{table}[t]
\centering\footnotesize
\caption{Possible 4-OPs obtained if a certain transcript is observed.}
\label{tabTrLag0}

\smallskip
\begin{tabular}{cccccc}
\toprule
$\tau_t=\pi^{[1]}$ & $\tau_t=\pi^{[2]}$ & $\tau_t=\pi^{[3]}$ & $\tau_t=\pi^{[4]}$ & $\tau_t=\pi^{[5]}$ & $\tau_t=\pi^{[6]}$ \\
 \midrule
$(1,2,3,4)$ & $(1,2,4,3)$ & $(2,1,3,4)$ & $(1,4,3,2)$ & $(1,3,2,4)$ & $(1,3,4,2)$ \\
$(4,3,2,1)$ & $(4,3,1,2)$ & $(3,4,2,1)$ & $(2,1,4,3)$ & $(1,4,2,3)$ & $(2,4,3,1)$ \\
 &  &  & $(2,4,1,3)$ & $(2,3,1,4)$ & $(3,1,2,4)$ \\
 &  &  & $(3,2,1,4)$ & $(2,3,4,1)$ & $(4,2,1,3)$ \\
 &  &  & $(3,2,4,1)$ & $(3,1,4,2)$ &  \\
 &  &  & $(4,1,3,2)$ & $(3,4,1,2)$ &  \\
 &  &  & $(4,2,3,1)$ & $(4,1,2,3)$ &  \\
 \bottomrule
 \end{tabular}
\end{table}

\medskip
Let $\fp_{\tau}$ be the stationary marginal distribution of the transcripts~$(\tau_t)$, \ie
\begin{equation*}
\fp_{\tau} = \big(p_{\tau;1}, \ldots, p_{\tau;6}\big)^\top :=\big(P(\tau_t=\pi^{[1]}), \ldots,
P(\tau_t)=\pi^{[6]}\big)^\top,
\end{equation*}
and $\fp_{\textup{K}}$ the stationary marginal distribution of the Kendall distances, \ie
\begin{equation*}
\fp_{\textup{K}} = (p_{\textup{K};0},\ldots,p_{\textup{K};3})^\top := \big(P( d_{\textup{K},t}=0), \ldots, P( d_{\textup{K},t}=3)\big)^\top. 
\end{equation*}

To calculate $\fp_{\textup{K}}$ from $\fp_{\tau}$, we need the following proposition \citep{weiss26}.

\begin{prop}
\label{PropDistancesCK}
It holds:
\begin{description}
\item[(a)] $d_{\textup{K}}(\pi _{1},\pi _{2})=0$ iff $\tau (\pi _{1},\pi
_{2})=\pi ^{[1]}$.

\item[(b)] $d_{\textup{K}}(\pi _{1},\pi _{2})=1$ iff $\tau (\pi _{1},\pi
_{2})\in \{\pi ^{[2]},\pi ^{[3]}\}$.

\item[(c)] $d_{\textup{K}}(\pi _{1},\pi _{2})=2$ iff $\tau (\pi _{1},\pi
_{2})\in \{\pi ^{[4]},\pi ^{[5]}\}$.

\item[(d)] $d_{\textup{K}}(\pi _{1},\pi _{2})=3$ iff $\tau (\pi _{1},\pi
_{2})=\pi ^{[6]}$.
\end{description}
\end{prop}
From Proposition~\ref{PropDistancesCK}, it follows that 
\ba
\label{trans_mat_K}
\fp_{\textup{K}}\ =\ \mT_{\textup{K}}\, \fp_{\tau},
\quad\text{where }
\mT_{\textup{K}}\ =\ \left(\begin{array}{cccccc}
1 & 0 & 0 & 0 & 0 & 0 \\
0 & 1 & 1 & 0 & 0 & 0 \\
0 & 0 & 0 & 1 & 1 & 0 \\
0 & 0 & 0 & 0 & 0 & 1 \\
\end{array}\right).
\ea
The following proposition \citep{weiss26} summarizes the relevant stochastic properties under the IC-assumption.

\begin{prop}
\label{prop_tr_marg_iid}
Let $(X_t)$ be \iid, so satisfying the IC-condition. Then, the transcripts' $(\tau_t)$ marginal distribution is given by
$$
    \fp_{\tau}^{(0)}\ =\ 
\tfrac{1}{24}\,\big(2, 2, 2, 7, 7, 4\big)^\top.
$$
Furthermore, the 
Kendall distances $(d_{\textup{K},t})$ have the IC-marginal distribution
$$
\begin{array}{@{}l}
\fp_{\textup{K}}^{(0)}\ =\ 
\tfrac{1}{12}\,\big(1, 2, 7, 2\big)^\top
\text{with mean } \mu_{\textup{K}}^{(0)}=\tfrac{11}{6}.
\end{array}
$$
\end{prop}
Note that the transcripts' IC-PMF, $\fp_{\tau}^{(0)}$, simply agrees with the relative frequencies of the respective 4-OPs in Table~\ref{tabTrLag0}.


\section{Novel Control Charts based on Transcripts}
\label{Novel Control Charts based on Transcripts}
Let $(X_t)$ be the continuously distributed and real-valued process that is assumed to be \iid\ under IC-conditions. Let $(\tau_t)$ be the corresponding series of transcripts, and $(d_{\textup{K},t})$ the series of Kendall distances. Recall from Proposition~\ref{prop_tr_marg_iid} that under IC-conditions, the transcripts' IC-distribution is given by $\fp_{\tau}^{(0)} = \tfrac{1}{24}\,\big(2, 2, 2, 7, 7, 4\big)^\top$, and the one of the Kendall distances by $\fp_{\textup{K}}^{(0)} = \tfrac{1}{12}\,\big(1, 2, 7, 2\big)^\top$ with mean $\mu_{\textup{K}}^{(0)}=\tfrac{11}{6}$. In analogy to equation \eqref{EWMA}, we first propose an EWMA-based online estimation with smoothing parameter $\lambda\in (0;1)$ of the transcripts' current PMF vector, namely via
\ba
\label{EWMA_tr}
\hat{\fp}_{\tau,1}\ =\ \fp_{\tau}^{(0)},\qquad
\hat{\fp}_{\tau,t}\ =\ \lambda\,\fZ_t + (1-\lambda)\,\hat{\fp}_{\tau,t-1} \quad\text{for } t=2,3,\ldots
\ea
Here, the binary vectors~$\fZ_t{}$ of dimension $m!=6$ are again defined by a ``one-hot encoding'', namely $Z_{t,k}=\indfkt(\tau_t=\pi^{[k]})$. The corresponding online estimate of $d_{\textup{K},t}$'s PMF and mean follow from 
\ba
\label{EWMA_dK}
\hat{\fp}_{\textup{K},t}\ =\ \mT_{\textup{K}}\, \hat{\fp}_{\tau,t}
\quad\text{and}\quad
\hat{\mu}_{\textup{K},t}\ =\ (0,1,2,3)\,\hat{\fp}_{\textup{K},t},
\ea
respectively, recall \eqref{trans_mat_K}. In view of the numerical experiments of \citet{weiss26} on hypothesis testing based on transcripts and algebraic distances, we now propose three novel types of control charts:

\begin{itemize}
	\item the $\Delta_\tau$-chart plots the statistics
\ba
\label{Delta_tau-chart}
\widehat{\Delta}_{\tau,t}\ =\ \sum_{k=1}^{m!} \frac{\big(\hat{p}_{\tau,t,k}-p_{\tau,k}^{(0)}\big)^2}{p_{\tau,k}^{(0)}}
\quad\text{for } t=2,3,\ldots
\ea
against an UCL $l_{\Delta_\tau}>0$;

	\item the $\Delta_{\textup{K}}$-chart plots the statistics
\ba
\label{Delta_K-chart}
\widehat{\Delta}_{\textup{K},t}\ =\ \sum_{i=0}^{3} \frac{\big(\hat{p}_{\textup{K},t,i}-p_{\textup{K},i}^{(0)}\big)^2}{p_{\textup{K},i}^{(0)}}
\quad\text{for } t=2,3,\ldots
\ea
against an UCL $l_{\Delta_{\textup{K}}}>0$;

  \item the $\mu_{\textup{K}}$-chart plots the statistics
\ba
\label{mu_K-chart}
\widehat{\mu}_{\textup{K},t}\ =\ \hat{\mu}_{\textup{K},t}-11/6
\quad\text{for } t=2,3,\ldots
\ea
against an UCL $l_{\mu_{\textup{K}}}>0$ and LCL $-l_{\mu_{\textup{K}}}$.
\end{itemize}
ARL computation and chart design of these charts are done by a simulation-based approach in analogy to \citet{weiss23}. Having specified a process model for $(X_t)$ as well as a value for~$\lambda$, one simulates the observations $X_1,X_2,\ldots$ and does an online computation of the considered chart statistic from \eqref{Delta_tau-chart}--\eqref{mu_K-chart}. The time when the statistic first violates a CL is the run length of the control chart. These run length simulations are repeated for sufficiently many times (in this research, we always use $10^5$ replications), and the sample mean across the replicated run lengths is an approximation of the control chart's true (zero-state) ARL. For computing an IC-ARL, we benefit from the nonparametric nature of charts \eqref{Delta_tau-chart}--\eqref{mu_K-chart}: as the marginal distribution is without effect on the charts' IC-performances, we can simply use a normal distribution for simulation. Then, the actual chart design is determined iteratively; having specified a target value for the IC-ARL (in accordance to the SPM literature, we use the target ARL$_0=370$), one sets an initial value of the CL-parameter and determines the corresponding IC-ARL. If the deviation to ARL$_0$ is too large (in this research, we always tried to keep the absolute amount of deviation below~1), then the CL-value is revised and the IC-ARL is updated, and this procedure is repeated until a satisfactory chart design is achieved. 


\section{Performance Analyses}
\label{Performance Analyses}
We analyze the performance of the newly proposed control charts \eqref{Delta_tau-chart}--\eqref{mu_K-chart} in comparison to the existing OP-EWMA charts surveyed in Section~\ref{Review: Control Charts based on Ordinal Patterns}, where we use the simulation-based approach described at the end of Section~\ref{Novel Control Charts based on Transcripts} for ARL calculation and chart design. As the first step, we compute the required chart designs (\ie the CLs of each chart) for achieving an IC-ARL close to ARL$_0=370$. As can be seen from Table~\ref{tab_IC-ARLs}, this was possible for any $\lambda\in\{0.25,0.10,0.05\}$ (ordered according to increasing memory), which are the same smoothing parameters as in Table~1 of \citet{weiss23}. Recall that the chart designs in Table~\ref{tab_IC-ARLs} apply to any continuously-distributed process $(X_t)$ due to the nonparametric nature of the charts \eqref{Delta_tau-chart}--\eqref{mu_K-chart}. They can be immediately used in practice, irrespective of the IC-model's marginal distribution and corresponding parametrization, and thus without the need for any parameter estimation.

\begin{table}[th]
\centering\small
\caption{Chart designs (CL~parameters $l_{\Delta_\tau},l_{\Delta_{\textup{K}}},l_{\mu_{\textup{K}}}$) for \eqref{Delta_tau-chart}--\eqref{mu_K-chart} and corresponding IC-ARLs for different values of smoothing parameter~$\lambda$.}
\label{tab_IC-ARLs}

\smallskip
\begin{tabular}{l|lr|lr|lr}
\toprule
 & \multicolumn{2}{l|}{$\Delta_\tau$-chart} & \multicolumn{2}{l|}{$\Delta_{\textup{K}}$-chart} & \multicolumn{2}{l}{$\mu_{\textup{K}}$-chart} \\
$\lambda$ & $l_{\Delta_\tau}$ & ARL & $l_{\Delta_{\textup{K}}}$ & ARL & $l_{\mu_{\textup{K}}}$ & ARL \\
\midrule
0.25 & 3.225 & 370.1 & 3.19 & 370.4 & 1.0188 & 369.8 \\
0.10 & 0.9685 & 369.8 & 0.8078 & 369.7 & 0.5827 & 370.3 \\
0.05 & 0.4328 & 370.2 & 0.3229 & 369.6 & 0.3785 & 370.5 \\
\bottomrule
\end{tabular}
\end{table}

\medskip
Next, we use these chart designs for evaluating the OOC-performance of our novel control charts \eqref{Delta_tau-chart}--\eqref{mu_K-chart} in comparison to the existing charts \eqref{Delta_pi-chart}--\eqref{beta-chart} of \citet{weiss23}. Therefore, we simulated OOC-ARLs for various types of serially dependent data-generating process (DGP), thus violating the IC-assumption of $(X_t)$ being \iid {} Here, we consider all OOC-DGPs that were already considered by \citet{weiss23}. We show the corresponding OOC-ARLs of charts \eqref{Delta_pi-chart}--\eqref{beta-chart} in italic font for comparison, but we also complement our analyses by an additional DGP from \citet{weiss26}, which is explained in more detail below.

\begin{table}[th]
\centering\scriptsize
\caption{OOC-ARLs of novel control charts \eqref{Delta_tau-chart}--\eqref{mu_K-chart} compared to existing charts \eqref{Delta_pi-chart}--\eqref{tau-chart} for AR$(1)$ DGP \eqref{AR1} with dependence parameter~$\alpha$ and smoothing parameter~$\lambda$. Italic numbers are taken from \citet{weiss23}. 
Lowest OOC-ARL among novel charts in bold font.}
\label{tab_OOC-ARLs_AR1}

\smallskip
\begin{tabular}{c|r@{\ \ }r|r@{\ \ }r@{\ \ }r|c|r@{\ \ }r|r@{\ \ }r@{\ \ }r}
\toprule
$\alpha$ & \multicolumn{1}{c}{$\Delta_\pi$} & \multicolumn{1}{c|}{$\uptau$} & \multicolumn{1}{c}{$\Delta_\tau$} & \multicolumn{1}{c}{$\Delta_{\textup{K}}$} & \multicolumn{1}{c|}{$\mu_{\textup{K}}$} & $\alpha$ & \multicolumn{1}{c}{$\Delta_\pi$} & \multicolumn{1}{c|}{$\uptau$} & \multicolumn{1}{c}{$\Delta_\tau$} & \multicolumn{1}{c}{$\Delta_{\textup{K}}$} & \multicolumn{1}{c}{$\mu_{\textup{K}}$} \\
\midrule
\multicolumn{12}{l}{$\lambda=0.25$} \\
\midrule
0.2 & \itshape 217.6 & \itshape 185.4 & 210.2 & 212.1 & \bfseries 181.1 & -0.2 & \itshape 640.2 & \itshape 798.2 & \bfseries 620.6 & 621.4 & 804.0 \\
0.4 & \itshape 128.5 & \itshape 101.8 & 116.7 & 118.4 & \bfseries 95.7 & -0.4 & \itshape 1162.0 & \itshape 1840.3 & 808.3 & \bfseries 795.1 & 1538.8 \\
0.6 & \itshape 75.3 & \itshape 60.4 & 68.6 & 68.7 & \bfseries 56.1 & -0.6 & \itshape 2266.1 & \itshape 4634.4 & 449.9 & \bfseries 432.2 & 969.6 \\
0.8 & \itshape 44.8 & \itshape 39.2 & 42.3 & 42.7 & \bfseries 35.6 & -0.8 & \itshape 5242.5 & \itshape 13649.7 & 116.0 & \bfseries 112.5 & 186.4 \\
\midrule
\multicolumn{12}{l}{$\lambda=0.10$} \\
\midrule
0.2 & \itshape 242.2 & \itshape 191.3 & 233.8 & 200.6 & \bfseries 171.4 & -0.2 & \itshape 531.3 & \itshape 411.2 & \bfseries 342.3 & 351.3 & 393.7 \\
0.4 & \itshape 152.9 & \itshape 99.9 & 123.4 & 101.1 & \bfseries 84.4 & -0.4 & \itshape 681.8 & \itshape 203.7 & 175.5 & \bfseries 153.4 & 155.8 \\
0.6 & \itshape 94.2 & \itshape 59.7 & 68.3 & 57.5 & \bfseries 49.5 & -0.6 & \itshape 565.1 & \itshape 83.8 & 72.6 & 60.2 & \bfseries 60.0 \\
0.8 & \itshape 57.6 & \itshape 40.5 & 42.4 & 36.3 & \bfseries 32.8 & -0.8 & \itshape 190.8 & \itshape 37.4 & 32.3 & \bfseries 27.6 & 28.7 \\
\midrule
\multicolumn{12}{l}{$\lambda=0.05$} \\
\midrule
0.2 & \itshape 269.5 & \itshape 206.2 & 233.7 & 181.4 & \bfseries 172.1 & -0.2 & \itshape 411.2 & \itshape 271.4 & 264.9 & 274.4 & \bfseries 246.5 \\
0.4 & \itshape 172.9 & \itshape 105.7 & 117.5 & 88.0 & \bfseries 82.4 & -0.4 & \itshape 326.2 & \itshape 118.0 & 119.3 & 104.5 & \bfseries 93.4 \\
0.6 & \itshape 108.7 & \itshape 63.8 & 66.4 & 51.6 & \bfseries 49.0 & -0.6 & \itshape 168.2 & \itshape 56.2 & 55.7 & 46.6 & \bfseries 44.3 \\
0.8 & \itshape 69.4 & \itshape 43.9 & 43.2 & 34.8 & \bfseries 33.7 & -0.8 & \itshape 71.3 & \itshape 30.7 & 30.0 & \bfseries 25.2 & 25.6 \\
\bottomrule
\end{tabular}
\end{table}

\medskip
The first OOC-DGP is the first-order autoregressive (AR$(1)$) process with \iid\ innovations $(\epsilon_t)$ being standard-normally distributed, 
\ba
\label{AR1}
X_t\ =\ \alpha\, X_{t-1} + \epsilon_t
\quad\text{with }
\epsilon_t\sim\norm(0,1),
\ea
see Table~\ref{tab_OOC-ARLs_AR1} for the obtained OOC-ARLs. Note that \citet{weiss23} omitted tabulating the OOC-ARLs of the $\upbeta$-chart \eqref{beta-chart}, because this statistic is not sensitive towards the linear dependence implied by the AR$(1)$ DGP. As AR$(1)$ dependence is probably the most common type of serial dependence in practice, it is highly satisfactory to see that our novel control charts (especially the $\Delta_{\textup{K}}$- and $\mu_{\textup{K}}$-chart) show an appealing OOC-performance regarding both positive and negative serial dependence. For positive dependence, we see that the value of~$\lambda$ often has an only mild effect on the OOC-ARLs, and there is also not much difference between the different types of control charts. The $\mu_{\textup{K}}$-chart is always the best choice (but $\Delta_{\tau}$ and $\Delta_{\textup{K}}$ perform similarly well), and it always outperforms any of the existing charts of \citet{weiss23}. The situation of negative dependence is much more sophisticated. Here, large values of~$\lambda$ lead to a poor chart performance, with OOC-ARLs being sometimes much larger than the IC-ARLs (``biased ARL profile''). The best ARL performance is achieved for $\lambda=0.05$, with again the $\mu_{\textup{K}}$-chart being the best overall solution. Hence, to sum up, the novel $\mu_{\textup{K}}$-chart clearly outperforms the existing OP-EWMA charts for both positive and negative AR$(1)$ dependence, provided that $\lambda= 0.05$ in the latter case.

\begin{table}[th]
\centering\small
\caption{OOC-ARLs of novel control charts \eqref{Delta_tau-chart}--\eqref{mu_K-chart} compared to existing charts \eqref{Delta_pi-chart}--\eqref{beta-chart} for TEAR$(1)$ DGP \eqref{TEAR1} with dependence parameter~$\alpha$ and smoothing parameter~$\lambda$. Italic numbers are taken from \citet{weiss23}. 
Lowest OOC-ARL among novel charts in bold font.}
\label{tab_OOC-ARLs_TEAR1}

\smallskip
\begin{tabular}{c|rrr|rrr}
\toprule
$\alpha$ & \multicolumn{1}{c}{$\Delta_\pi$} & \multicolumn{1}{c}{$\upbeta$} & \multicolumn{1}{c|}{$\uptau$} & \multicolumn{1}{c}{$\Delta_\tau$} & \multicolumn{1}{c}{$\Delta_{\textup{K}}$} & \multicolumn{1}{c}{$\mu_{\textup{K}}$} \\
\midrule
\multicolumn{7}{l}{$\lambda=0.25$} \\
\midrule
0.1 & \itshape 252.0 & \itshape 254.6 & \itshape 265.8 & 264.6 & 267.0 & \bfseries 257.8 \\
0.2 & \itshape 135.3 & \itshape 133.7 & \itshape 166.5 & \bfseries 156.5 & 158.2 & 157.8 \\
0.3 & \itshape 70.9 & \itshape 69.1 & \itshape 96.3 & \bfseries 86.1 & 87.2 & 90.4 \\
0.4 & \itshape 39.8 & \itshape 38.5 & \itshape 55.8 & \bfseries 49.1 & 49.9 & 51.9 \\
0.5 & \itshape 24.1 & \itshape 23.4 & \itshape 33.0 & \bfseries 29.3 & 29.8 & 31.0 \\
0.6 & \itshape 15.6 & \itshape 15.3 & \itshape 20.5 & \bfseries 18.4 & 18.7 & 19.3 \\
\midrule
\multicolumn{7}{l}{$\lambda=0.10$} \\
\midrule
0.1 & \itshape 238.4 & \itshape 231.1 & \itshape 286.0 & \bfseries 271.9 & 276.6 & 272.5 \\
0.2 & \itshape 115.5 & \itshape 108.8 & \itshape 191.0 & \bfseries 155.5 & 169.9 & 175.0 \\
0.3 & \itshape 59.2 & \itshape 55.1 & \itshape 113.6 & \bfseries 82.7 & 93.3 & 102.8 \\
0.4 & \itshape 34.4 & \itshape 32.0 & \itshape 65.1 & \bfseries 46.7 & 51.9 & 59.3 \\
0.5 & \itshape 22.1 & \itshape 20.6 & \itshape 38.3 & \bfseries 28.5 & 30.6 & 35.9 \\
0.6 & \itshape 15.3 & \itshape 14.5 & \itshape 23.6 & \bfseries 18.7 & 19.2 & 22.4 \\
\midrule
\multicolumn{7}{l}{$\lambda=0.05$} \\
\midrule
0.1 & \itshape 230.4 & \itshape 214.4 & \itshape 310.7 & \bfseries 265.3 & 284.0 & 290.4 \\
0.2 & \itshape 107.7 & \itshape 95.7 & \itshape 213.7 & \bfseries 143.5 & 177.0 & 188.5 \\
0.3 & \itshape 57.3 & \itshape 50.6 & \itshape 126.9 & \bfseries 77.2 & 97.9 & 110.2 \\
0.4 & \itshape 35.1 & \itshape 30.9 & \itshape 72.5 & \bfseries 45.9 & 54.9 & 64.2 \\
0.5 & \itshape 23.9 & \itshape 21.2 & \itshape 43.3 & \bfseries 29.8 & 32.8 & 39.2 \\
0.6 & \itshape 17.4 & \itshape 15.5 & \itshape 27.1 & \bfseries 20.6 & 20.9 & 25.3 \\
\bottomrule
\end{tabular}
\end{table}

\smallskip
Next, we consider the transposed exponential AR$(1)$ (TEAR$(1)$) process with standard-exponential \iid\ innovations $(\varepsilon_t)$, defined by
\ba
\label{TEAR1}
X_t\ =\ B_t^{(\alpha)}\, X_{t-1} + (1-\alpha)\,\varepsilon_t
\quad\text{with }
\varepsilon_t\sim\expon(1),
\ea
where $(B_t^{(\alpha)})$ are \iid\ Bernoulli random variables with $P(B_t^{(\alpha)}=1)=\alpha$. The obtained OOC-ARLs are summarized in Table~\ref{tab_OOC-ARLs_TEAR1}. The TEAR$(1)$ DGP is quite demanding in a sense. On the one hand, its autocorrelation function (ACF) is exponentially decaying like in the AR$(1)$ case, so it may be classified as being ``linear'' in this respect. On the other hand, it behaves highly asymmetric as the generated sample paths are characterized by long-lasting rises that are interupted by abrupt falls. Recalling the definition of the 3-OPs in \eqref{OrdPatt3}, it is clear that OP $\pi^{[1]}=(1,2,3)$ occurs much more frequently than $\pi^{[6]}=(3,2,1)$ (and also than the other OPs). Thus, the $\upbeta$-chart \eqref{beta-chart} is tailor-made for detecting this kind of dependence, which explains its superior performance in the analyses of \citet{weiss23}. From Table~\ref{tab_OOC-ARLs_TEAR1}, we recognize that our novel control charts \eqref{Delta_tau-chart}--\eqref{mu_K-chart} do not reach this outstanding performance (with the $\Delta_\tau$-chart being the best among them), but they are at least competitive and better than the $\uptau$-chart of \citet{weiss23}.

\begin{table}[th]
\centering\small
\caption{OOC-ARLs of novel control charts \eqref{Delta_tau-chart}--\eqref{mu_K-chart} compared to existing charts \eqref{Delta_pi-chart}--\eqref{beta-chart} for AAR$(1)$ DGP \eqref{AAR1} with dependence parameter~$\alpha$ and smoothing parameter~$\lambda$. Italic numbers are taken from \citet{weiss23}. 
Lowest OOC-ARL among novel charts in bold font.}
\label{tab_OOC-ARLs_AAR1}

\smallskip
\begin{tabular}{c|rrr|rrr}
\toprule
$\alpha$ & \multicolumn{1}{c}{$\Delta_\pi$} & \multicolumn{1}{c}{$\upbeta$} & \multicolumn{1}{c|}{$\uptau$} & \multicolumn{1}{c}{$\Delta_\tau$} & \multicolumn{1}{c}{$\Delta_{\textup{K}}$} & \multicolumn{1}{c}{$\mu_{\textup{K}}$} \\
\midrule
\multicolumn{7}{l}{$\lambda=0.25$} \\
\midrule
0.2 & \itshape 329.6 & \itshape 334.0 & \itshape 326.6 & 330.5 & 333.0 & \bfseries 323.3 \\
0.4 & \itshape 237.8 & \itshape 252.9 & \itshape 229.0 & 240.9 & 243.2 & \bfseries 222.7 \\
0.6 & \itshape 147.0 & \itshape 169.9 & \itshape 132.0 & 145.7 & 147.3 & \bfseries 125.4 \\
0.8 & \itshape 79.3 & \itshape 101.7 & \itshape 68.1 & 74.6 & 75.4 & \bfseries 63.0 \\
\midrule
\multicolumn{7}{l}{$\lambda=0.10$} \\
\midrule
0.2 & \itshape 322.7 & \itshape 323.0 & \itshape 334.3 & 329.8 & 336.6 & \bfseries 329.3 \\
0.4 & \itshape 233.2 & \itshape 238.1 & \itshape 242.5 & 245.2 & 243.9 & \bfseries 224.4 \\
0.6 & \itshape 154.0 & \itshape 170.2 & \itshape 138.2 & 149.2 & 135.9 & \bfseries 119.5 \\
0.8 & \itshape 93.5 & \itshape 123.5 & \itshape 70.7 & 76.1 & 65.8 & \bfseries 58.1 \\
\midrule
\multicolumn{7}{l}{$\lambda=0.05$} \\
\midrule
0.2 & \itshape 320.5 & \itshape 313.1 & \itshape 346.5 & \bfseries 323.3 & 338.4 & 340.1 \\
0.4 & \itshape 234.1 & \itshape 224.7 & \itshape 264.3 & \bfseries 232.1 & 239.0 & 235.7 \\
0.6 & \itshape 160.7 & \itshape 164.9 & \itshape 150.4 & 141.5 & 126.2 & \bfseries 121.4 \\
0.8 & \itshape 104.7 & \itshape 136.6 & \itshape 76.0 & 74.6 & 61.0 & \bfseries 58.7 \\
\bottomrule
\end{tabular}
\end{table}

\smallskip
The third OOC-DGP is the absolute AR$(1)$ (AAR$(1)$) process defined by 
\ba
\label{AAR1}
X_t\ =\ \alpha\, |X_{t-1}| + \epsilon_t
\quad\text{with }
\epsilon_t\sim\norm(0,1),
\ea
see Table~\ref{tab_OOC-ARLs_AAR1} for the obtained OOC-ARLs. The process is nonlinear with an again asymmetric behavior, as the absolute value~$|X_{t-1}|$ only leads to positive contributions. From Table~\ref{tab_OOC-ARLs_AAR1}, we recognize that the choice of~$\lambda$ is without a strong effect and that there is not much difference among the different control charts, similar to the classical AR$(1)$ process \eqref{AR1} with positive~$\alpha$. In most cases, the $\mu_{\textup{K}}$-chart is superior among the novel charts, and it usually also outperforms all existing OP-EWMA charts.

\begin{table}[th]
\centering\small
\caption{OOC-ARLs of novel control charts \eqref{Delta_tau-chart}--\eqref{mu_K-chart} compared to existing charts \eqref{Delta_pi-chart}--\eqref{beta-chart} for QAR$(1)$ DGP \eqref{QAR1} with dependence parameter~$\alpha$ and smoothing parameter~$\lambda$. Italic numbers are taken from \citet{weiss23}. 
Lowest OOC-ARL among novel charts in bold font.}
\label{tab_OOC-ARLs_QAR1}

\smallskip
\begin{tabular}{c|rrr|rrr}
\toprule
$\alpha$ & \multicolumn{1}{c}{$\Delta_\pi$} & \multicolumn{1}{c}{$\upbeta$} & \multicolumn{1}{c|}{$\uptau$} & \multicolumn{1}{c}{$\Delta_\tau$} & \multicolumn{1}{c}{$\Delta_{\textup{K}}$} & \multicolumn{1}{c}{$\mu_{\textup{K}}$} \\
\midrule
\multicolumn{7}{l}{$\lambda=0.25$} \\
\midrule
0.15 & \itshape 283.8 & \itshape 290.3 & \itshape 285.1 & 290.5 & 292.9 & \bfseries 282.0 \\
0.2 & \itshape 219.3 & \itshape 230.5 & \itshape 217.2 & 223.7 & 225.8 & \bfseries 213.3 \\
0.25 & \itshape 116.0 & \itshape 123.8 & \itshape 114.9 & 117.5 & 118.2 & \bfseries 110.6 \\
0.3 & \itshape 35.7 & \itshape 37.6 & \itshape 35.5 & 36.0 & 36.2 & \bfseries 34.7 \\
\midrule
\multicolumn{7}{l}{$\lambda=0.10$} \\
\midrule
0.15 & \itshape 273.6 & \itshape 271.7 & \itshape 302.8 & \bfseries 286.6 & 297.2 & 288.4 \\
0.2 & \itshape 214.7 & \itshape 215.1 & \itshape 234.0 & 224.8 & 228.4 & \bfseries 218.8 \\
0.25 & \itshape 116.6 & \itshape 118.6 & \itshape 122.3 & 119.8 & 118.6 & \bfseries 113.2 \\
0.3 & \itshape 37.3 & \itshape 37.9 & \itshape 37.7 & 37.6 & 36.3 & \bfseries 35.9 \\
\midrule
\multicolumn{7}{l}{$\lambda=0.05$} \\
\midrule
0.15 & \itshape 268.8 & \itshape 254.3 & \itshape 321.9 & \bfseries 271.2 & 299.1 & 303.1 \\
0.2 & \itshape 213.6 & \itshape 200.4 & \itshape 257.8 & \bfseries 212.7 & 230.0 & 232.3 \\
0.25 & \itshape 120.0 & \itshape 113.6 & \itshape 134.3 & \bfseries 117.2 & 119.8 & 120.4 \\
0.3 & \itshape 39.5 & \itshape 38.3 & \itshape 40.7 & 38.6 & \bfseries 37.5 & 38.0 \\
\bottomrule
\end{tabular}
\end{table}

\smallskip
A related OOC-DGP is the quadratic AR$(1)$ (QAR$(1)$) process defined by 
\ba
\label{QAR1}
X_t\ =\ \alpha\, X_{t-1}^2 + \epsilon_t
\quad\text{with }
\epsilon_t\sim\norm(0,1),
\ea
see Table~\ref{tab_OOC-ARLs_QAR1}, which is characterized by an analogous asymmetry as for the AAR$(1)$ process \eqref{AAR1}, but with more pronounced positive shocks due to the squared term~$X_{t-1}^2$. Altogether, the QAR$(1)$ process is highly nonlinear, but all charts from Table~\ref{tab_OOC-ARLs_AAR1} are well-suited for detecting such kind of dependence. As before, we recognize only moderate differences between the different charts and different $\lambda$-values, where the $\mu_{\textup{K}}$-chart for $\lambda\geq 0.10$ and the $\Delta_\tau$-chart for $\lambda=0.05$ are best among the novel control charts. They often also outperform the best OP-EWMA chart, or perform at least very similar to it.

\begin{table}[th]
\centering\small
\caption{OOC-ARLs of novel control charts \eqref{Delta_tau-chart}--\eqref{mu_K-chart} compared to existing charts \eqref{Delta_pi-chart}--\eqref{beta-chart} for QMA$(1)$ DGP \eqref{QMA1} with dependence parameter~$\beta$ and smoothing parameter~$\lambda$. 
Lowest OOC-ARL among novel charts in bold font.}
\label{tab_OOC-ARLs_QMA1}

\smallskip
\begin{tabular}{c|rrr|rrr}
\toprule
$\beta$ & \multicolumn{1}{c}{$\Delta_\pi$} & \multicolumn{1}{c}{$\upbeta$} & \multicolumn{1}{c|}{$\uptau$} & \multicolumn{1}{c}{$\Delta_\tau$} & \multicolumn{1}{c}{$\Delta_{\textup{K}}$} & \multicolumn{1}{c}{$\mu_{\textup{K}}$} \\
\midrule
\multicolumn{7}{l}{$\lambda=0.25$} \\
\midrule
0.2 & 310.3 & 310.1 & 321.8 & 324.3 & 327.1 & \bfseries 323.5 \\
0.4 & 250.7 & 251.9 & 263.2 & 267.8 & 270.7 & \bfseries 261.4 \\
0.6 & 223.0 & 227.2 & 230.1 & 235.8 & 238.2 & \bfseries 228.5 \\
0.8 & 213.6 & 219.1 & 219.3 & 223.6 & 226.1 & \bfseries 217.0 \\
\midrule
\multicolumn{7}{l}{$\lambda=0.10$} \\
\midrule
0.2 & 277.5 & 272.9 & 334.6 & \bfseries 286.3 & 317.4 & 323.6 \\
0.4 & 213.2 & 207.2 & 281.1 & \bfseries 215.1 & 253.4 & 261.1 \\
0.6 & 200.7 & 193.3 & 248.3 & \bfseries 192.8 & 219.8 & 228.7 \\
0.8 & 204.4 & 195.0 & 238.5 & \bfseries 190.5 & 208.8 & 219.3 \\
\midrule
\multicolumn{7}{l}{$\lambda=0.05$} \\
\midrule
0.2 & 257.1 & 244.3 & 349.1 & \bfseries 253.3 & 321.0 & 331.1 \\
0.4 & 194.2 & 179.0 & 303.8 & \bfseries 177.5 & 256.9 & 275.0 \\
0.6 & 190.1 & 171.9 & 271.3 & \bfseries 163.8 & 224.5 & 243.6 \\
0.8 & 201.0 & 180.3 & 262.3 & \bfseries 168.5 & 213.9 & 234.7 \\
\bottomrule
\end{tabular}
\end{table}

\smallskip
While \citet{weiss23} restricted their analyses to solely AR-type processes, we follow \citet{weiss26} and also consider the first-order quadratic moving-average (QMA$(1)$) process, defined by 
\ba
\label{QMA1}
X_t\ =\ \epsilon_t + \beta\, \epsilon_{t-1}^2
\quad\text{with }
\epsilon_t\sim\norm(0,1),
\ea
as a further OOC-DGP. The (highly nonlinear and asymmetric) QMA$(1)$ process is 1-dependent, which means that~$X_t$ and~$X_{t-h}$ are independent of each other for time lags $h\geq 2$. The rather large OOC-ARLs in Table~\ref{tab_OOC-ARLs_QMA1} show that QMA$(1)$~dependence is generally difficult to detect, where the $\Delta_\tau$-chart together with $\lambda=0.05$ has the clearly best ARL performance among all novel charts. Its main competitor is the $\upbeta$-chart of \citet{weiss23}, also with $\lambda=0.05$, which has a slightly lower OOC-ARL for the lowest dependence level $\beta=0.2$, but is outperformed by the $\Delta_\tau$-chart otherwise.

\medskip
In a nutshell, the newly proposed control charts \eqref{Delta_tau-chart}--\eqref{mu_K-chart}, which are based on transcripts and algebraic distances, show an appealing ARL~performance compared to the existing OP-EWMA charts \eqref{Delta_pi-chart}--\eqref{beta-chart} of \citet{weiss23}. In the case of the positively dependent AR-type processes, there is not much difference between the charts \eqref{Delta_tau-chart}--\eqref{mu_K-chart}, and also the actual choice of~$\lambda$ has only little effect. These properties are attractive for the practitioner, as there is no ``bad choice'' regarding these types of dependence structure. If, in turn, one is faced with possible negative AR$(1)$~dependence or QMA$(1)$~dependence, then the small smoothing parameter $\lambda=0.05$ is clearly recommended, together with the $\mu_{\textup{K}}$-chart in the first case, and the $\Delta_\tau$-chart in the second case. Generally, one of these two charts, $\mu_{\textup{K}}$ or $\Delta_\tau$, usually performed best among the novel charts \eqref{Delta_tau-chart}--\eqref{mu_K-chart}, and except for the TEAR$(1)$ process with its exceptional sample paths, they usually also outperform the OP-EWMA charts \eqref{Delta_pi-chart}--\eqref{beta-chart}.

\begin{figure}[t]
\centering\scriptsize
(a)\hspace{-3ex}\includegraphics[viewport=0 45 405 270, clip=, scale=0.5]{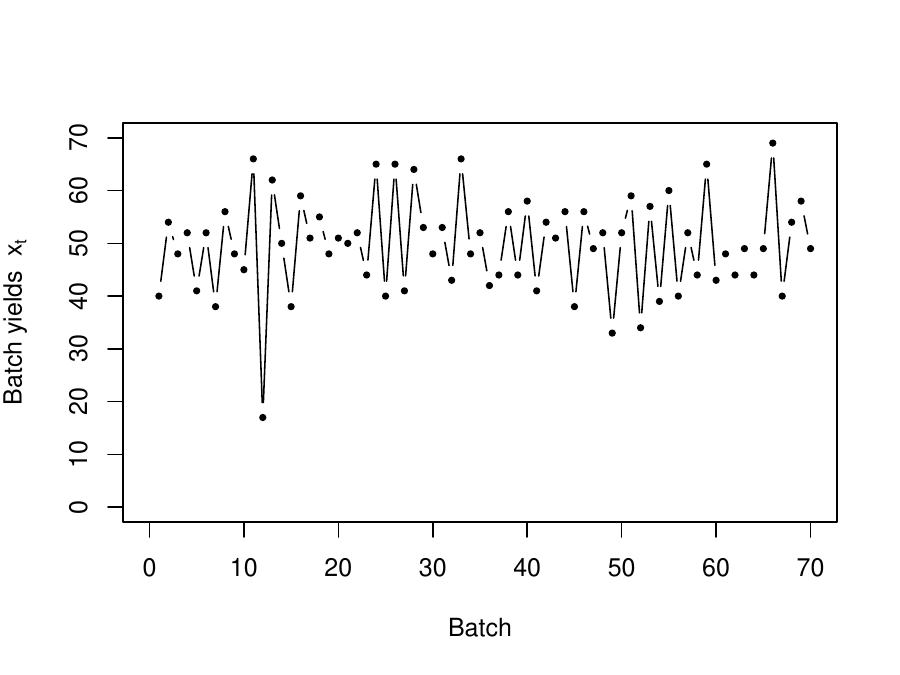}\!$t$
\quad
(b)\hspace{-3ex}\includegraphics[viewport=0 45 260 270, clip=, scale=0.5]{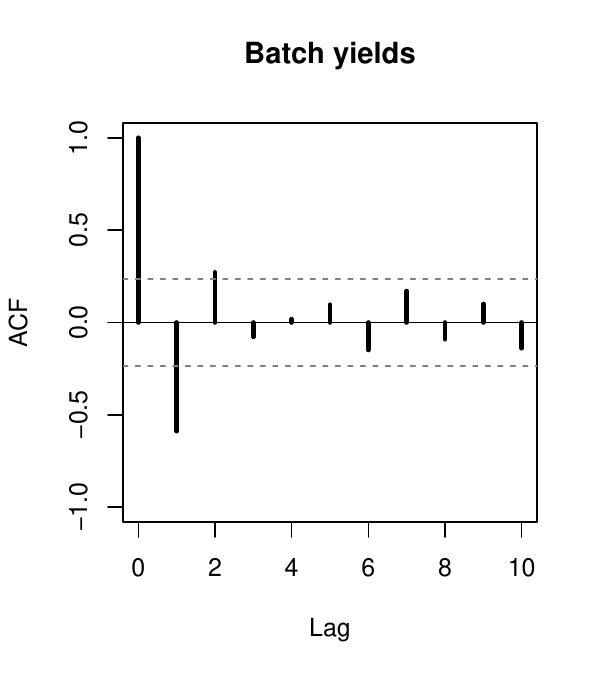}\,$h$
\caption{Chemical process data $(x_t)$ of Section~\ref{Illustrative Data Application}: (a) time series plot, and (b) plot of sample ACF against time lag~$h$.}
\label{figChemProcess}
\end{figure}

\section{Illustrative Data Application}
\label{Illustrative Data Application}
Inspired by the industrial application discussed in \citet{weiss23}, let us analyze the data set printed in Appendix A.3 of \citet[``Time series~4.1'']{odonovan83}, where the $n = 70$ consecutive yields from a batch chemical process as plotted in Figure~\ref{figChemProcess}\,(a) are provided. Since our novel transcript-based charts \eqref{Delta_tau-chart}--\eqref{mu_K-chart} are nonparametric (similar to the former OP-EWMA charts of \citet{weiss23}), they can immediately be applied to these data (using the chart designs from Table~\ref{tab_IC-ARLs}) without the need for any model fitting, in order to monitor the IC-assumptions of the batch yields being \iid\ (irrespective of the actual marginal distribution). However, we can expect this IC-assumption to be violated, because batch data are known to commonly exhibit negative values for the ACF. According to \citet[p.~34]{odonovan83}, high-yielding batches often cause residues that reduce the yield of the respective subsequent batch, which explains the alternating behavior being visible in Figure~\ref{figChemProcess}\,(a). In fact, the plotted sample ACF in Figure~\ref{figChemProcess}\,(b) confirms this conjecture with a strongly negative lag-1 ACF of about~$-0.588$. Therefore, the OOC-ARL performance according to the right part of Table~\ref{tab_OOC-ARLs_AR1} (lines ``$\alpha=-0.6$'') appears to be a reasonable benchmark for interpreting the subsequent control charts.

\begin{figure}[th!]
\centering\scriptsize
(a)\hspace{-3ex}\includegraphics[viewport=0 45 335 270, clip=, scale=0.5]{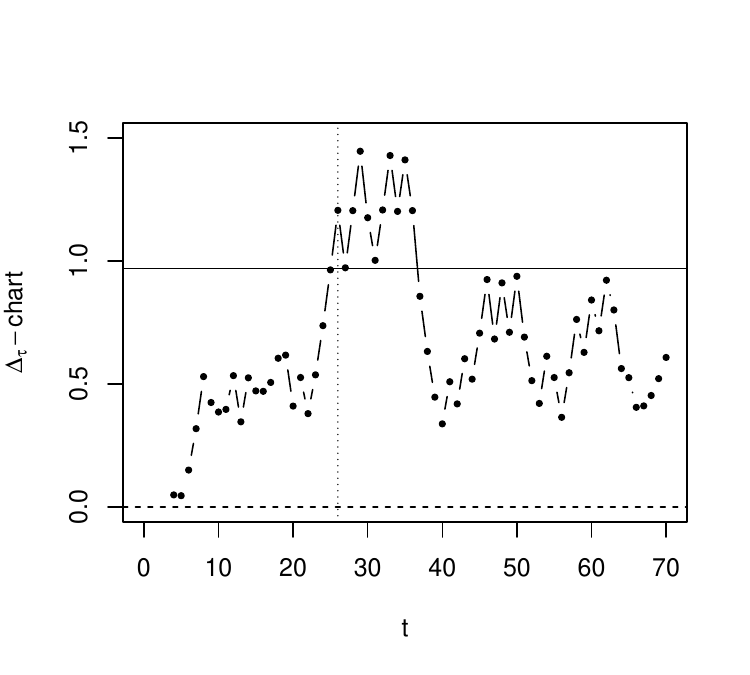}\!$t$
\quad
(b)\hspace{-3ex}\includegraphics[viewport=0 45 335 270, clip=, scale=0.5]{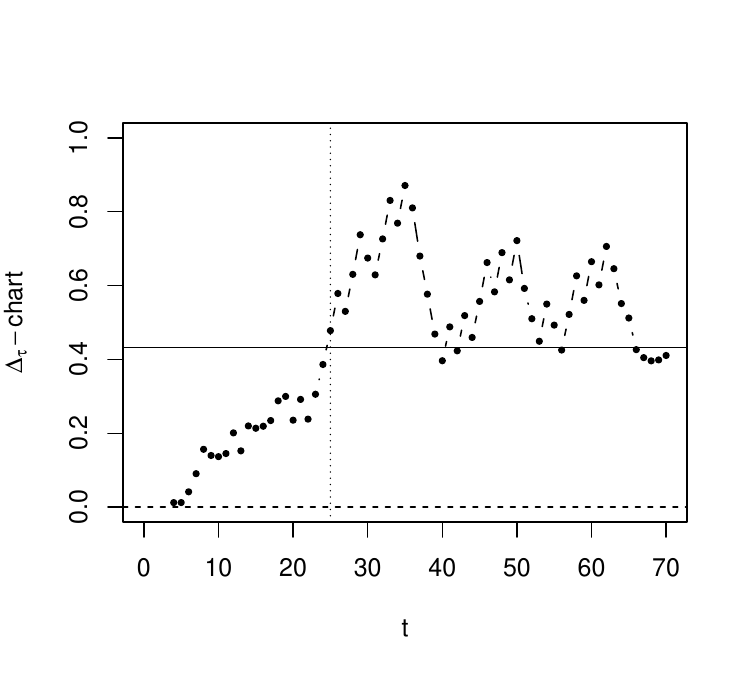}\!$t$
\\[2ex]
(c)\hspace{-3ex}\includegraphics[viewport=0 45 335 270, clip=, scale=0.5]{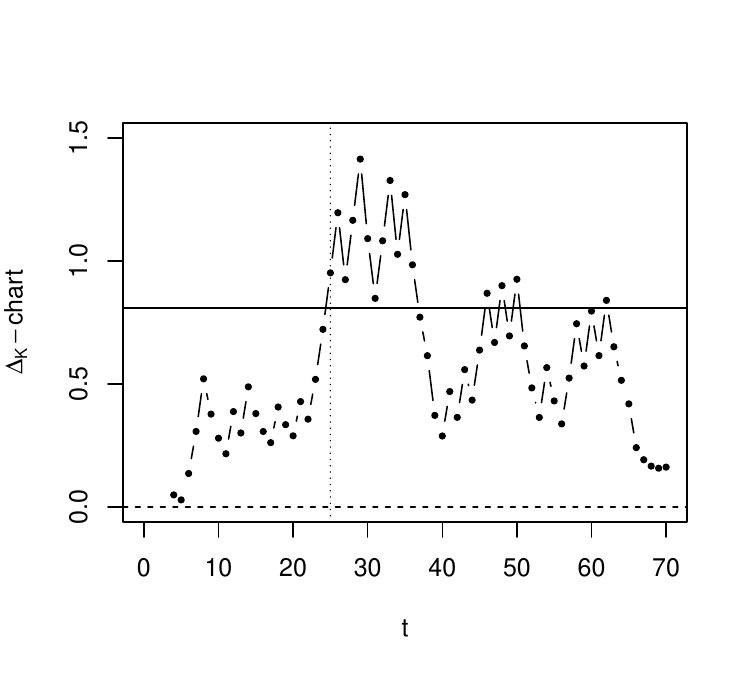}\!$t$
\quad
(d)\hspace{-3ex}\includegraphics[viewport=0 45 335 270, clip=, scale=0.5]{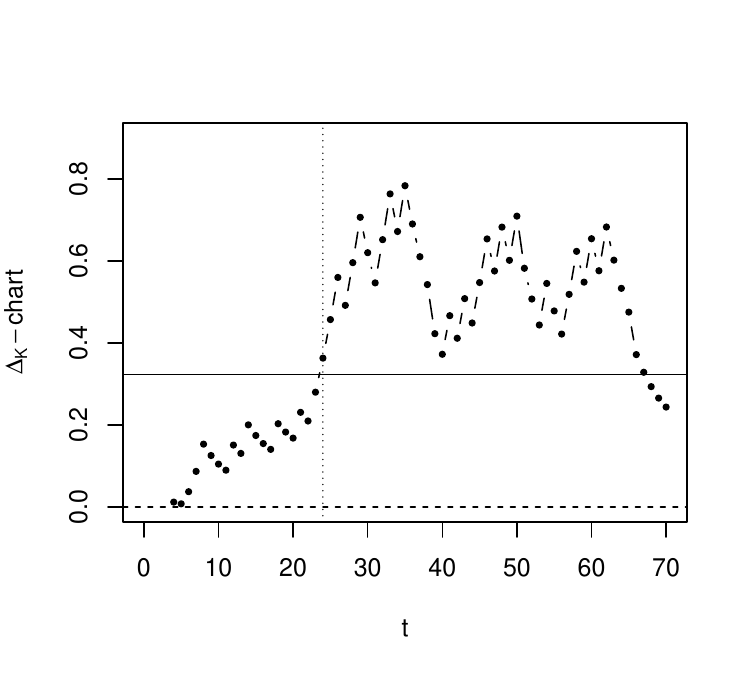}\!$t$
\\[2ex]
(e)\hspace{-3ex}\includegraphics[viewport=0 45 335 270, clip=, scale=0.5]{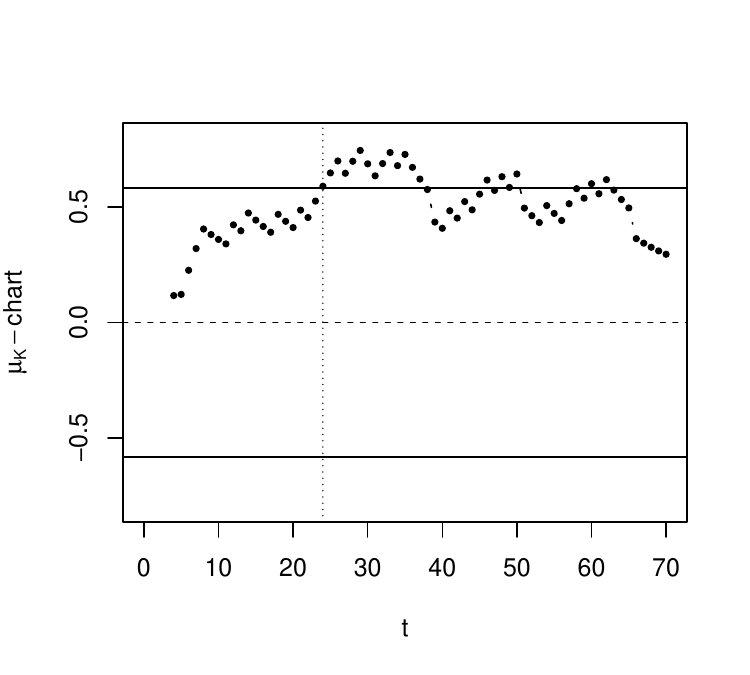}\!$t$
\quad
(f)\hspace{-3ex}\includegraphics[viewport=0 45 335 270, clip=, scale=0.5]{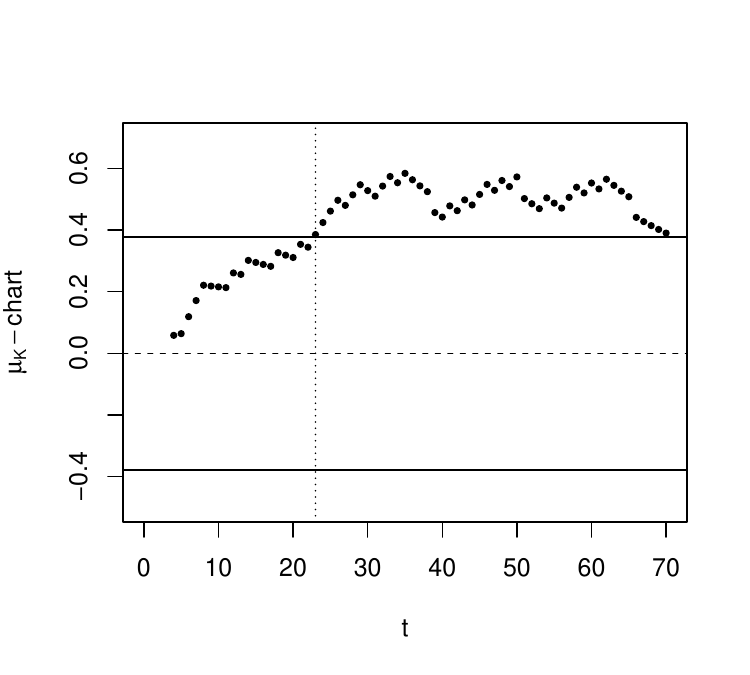}\!$t$
\caption{Chemical process data $(x_t)$ of Section~\ref{Illustrative Data Application}: $\Delta_\tau$-, $\Delta_{\textup{K}}$-, and $\mu_{\textup{K}}$-charts for $\lambda=0.10$ (left column) and $\lambda=0.05$ (right column).}
\label{figChemProcessCharts}
\end{figure}

\medskip
We first applied our novel control charts \eqref{Delta_tau-chart}--\eqref{mu_K-chart} with $\lambda=0.25$, but did not receive any alarm. This is plausible in view of the aforementioned Table~\ref{tab_OOC-ARLs_AR1}, where the charts are not sensitive to negative dependence for such a large value of~$\lambda$. However, we would expect a better OOC-performance if using $\lambda\leq 0.10$. This is confirmed by Figure~\ref{figChemProcessCharts}, where all charts signal a violation of the IC-assumption. In fact, in agreement to Table~\ref{tab_OOC-ARLs_AR1}, the charts with $\lambda=0.05$ are faster in triggering their first alarm than those with $\lambda=0.10$, and the $\mu_{\textup{K}}$-chart \eqref{mu_K-chart} is faster than the $\Delta_{\textup{K}}$-chart \eqref{Delta_K-chart} and the $\Delta_\tau$-chart \eqref{Delta_tau-chart}. The exact alarm times are
\begin{itemize}
    \item $\lambda=0.10$:\quad $t=26$ for $\Delta_\tau$, $t=25$ for $\Delta_{\textup{K}}$, and $t=24$ for $\mu_{\textup{K}}$;
    \item $\lambda=0.05$:\quad $t=25$ for $\Delta_\tau$, $t=24$ for $\Delta_{\textup{K}}$, and $t=23$ for $\mu_{\textup{K}}$.
\end{itemize}
Note that we express the alarm times in terms of the original sampling time according to Figure~\ref{figChemProcessCharts}, in order to allow for a fair judgement: for computing the first $k$~transcripts, we indeed have to collect $k+3$ original observations, because the $k$th transcript is computed from the segment $(x_k,\ldots,x_{k+3})$.

\smallskip
Finally, we also applied the OP-EWMA charts from Section~\ref{Review: Control Charts based on Ordinal Patterns} for comparison. More precisely, as already explained for Table~\ref{tab_OOC-ARLs_AR1}, the $\upbeta$-chart is not useful for linear dependence, so we focus on the OP-EWMA charts \eqref{Delta_pi-chart} and \eqref{tau-chart} of \citet{weiss23} as the competitors. Here, the $\Delta_\pi$-chart \eqref{Delta_pi-chart} does not trigger an alarm at all for $\lambda\geq 0.10$, and only at $t=62$ for $\lambda=0.05$. This poor performance appears reasonable in view of Table~\ref{tab_OOC-ARLs_AR1}. It is also reasonable that the $\uptau$-chart \eqref{tau-chart} is much faster, namely $t=26$ if $\lambda=0.10$ and $t=24$ if $\lambda=0.05$. But altogether, $\uptau$ is outperformed by~$\Delta_{\textup{K}}$ and~$\mu_{\textup{K}}$, in agreement to our findings from Table~\ref{tab_OOC-ARLs_AR1}. So we confirm the appealing performance of our novel transcript-based control charts.


\section{Conclusions}
\label{Conclusions}

The topic of this article is the application of transcripts to control
charts, \ie to the detection of change points regarding serial dependence in (continuously distributed) stochastic processes $(X_{t})$. The context is the analysis of time
series in ordinal representations, meaning symbolic representations whose
symbols (OPs) can be
viewed as permutations. In comparison to \citet{weiss23}, where OPs were used
to the same end, here, we go further and exploit two important features of permutations.

\begin{description}
\item[(1)] \textbf{The algebraic structure}. Permutations on $m$ objects,
endowed with function composition ``$\circ $'', build the symmetric group $S_{m}$ of
degree $m$. This algebraic structure allows to define the concept of
transcript from $\pi _{1}\in S_{m}$ to $\pi _{2}\in S_{m}$ as $\tau (\pi
_{1},\pi _{2})=\pi _{2}\circ \pi _{1}^{-1}\in S_{m}$, which is a
generalization of the concept of difference in additive groups. In this
paper, we focus on the case $m=3$.

\item[(2)] \textbf{The metric structure}. Permutations on $m$ objects build
a metric space, too. In Section~\ref{Review: Transcripts and Algebraic Distances}, we present two algebraic distances in $%
S_{m}$: the Cayley distance $d_{C}$ and the Kendall distance $d_{K}$. As it
turns out, both $d_{C}(\pi _{1},\pi _{2})$ and $d_{K}(\pi _{1},\pi _{2})$
can be easily calculated from the transcript $\tau (\pi _{1},\pi _{2})$; see
equations (12) and (13). In this paper, we only use the Kendall distance due to its better performance in previous works.
\end{description}
With these two tools of ordinal representations at hand --- transcripts and
transcript-based distances --- we proceed as follows. The real-valued process $(X_{t})$ is converted (in an online manner) into, first,
an OP-valued process $(\pi _{t})$ (Section \ref{Review: Control Charts based on Ordinal Patterns}), and second, into
the OP-valued process $(\tau _{t})=(\tau (\pi _{t},\pi _{t+1}))$. Then,
one monitors statistical properties (in particular, serial dependence) of
the process $(X_{t})$ via $(\tau _{t})$ and/or the integer-valued process $(d_{K,t})=(d_{K}(\pi _{t},\pi _{t+1}))$.

\medskip
We apply the above approach to control charts in Sections~\ref{Novel Control Charts based on Transcripts}--\ref{Illustrative Data Application}, the in-control assumption (``null hypothesis'') being that $(X_{t})$ is \iid {} To monitor for dependence changes in $(X_t)$, in Section~\ref{Novel Control Charts based on Transcripts}, we propose three novel, transcript-based and nonparametric control charts: the $\Delta _{\tau }$-chart \eqref{Delta_tau-chart}, the $\Delta _{\textup{K}}$-chart \eqref{Delta_K-chart}, and the $\mu _{\textup{K}}$-chart \eqref{mu_K-chart}. Their ARL performance is compared to that of the existing OP-EWMA charts \eqref{Delta_pi-chart}--\eqref{beta-chart} using different serially dependent processes (equations
\eqref{AR1}--\eqref{QMA1}) in Section~\ref{Performance Analyses}. The same methodology is applied in
Section~\ref{Illustrative Data Application}, this time to real-world data. There, consecutive yields from
a batch chemical process are monitored, with batch data being known to generally exhibit
negative values of the ACF. The numerical simulations in Section~\ref{Performance Analyses}
show a very satisfactory performance of the novel nonparametric control charts. Altogether, the best
performers are the charts $\mu _{\textup{K}}$ and $\Delta _{\tau }$. These positive outcomes are also confirmed by the
chemical process data in
Section~\ref{Illustrative Data Application}, with $\Delta _{\textup{K}}$ and $\mu _{\textup{K}}$ now offering the best
performance. In view of the above encouraging results, we conclude that the transcript-based control charts $\Delta _{\tau}$, $\Delta _{\textup{K}}$, and $\mu _{\textup{K}}$ are useful for a nonparametric monitoring of serial dependence.

\medskip
Future research will explore the possibility of extending our transcript-based approach to monitoring spatial OPs like in \citet{adaemmer26}.







%
%
%

\end{document}